Astronomy
&
Astrophysics

LETTER TO THE EDITOR

# Strong H$\alpha$ emission and signs of accretion in a circumbinary planetary mass companion from MUSE

Simon C. Eriksson[1], Rubén Asensio Torres[2], Markus Janson[1], Yuhiko Aoyama[3,4], Gabriel-Dominique Marleau[5,6,2], Mickael Bonnefoy[7], and Simon Petrus[7]

[1] Institutionen för astronomi, Stockholms universitet, AlbaNova universitetscentrum, 106 91 Stockholm, Sweden
  e-mail: simon.eriksson@astro.su.se
[2] Max-Planck-Institut für Astronomie, Königstuhl 17, 69117 Heidelberg, Germany
[3] Institute for Advanced Study, Tsinghua University, Beijing 100084, PR China
[4] Department of Astronomy, Tsinghua University, Beijing 100084, PR China
[5] Institut für Astronomie und Astrophysik, Tübingen, Auf der Morgenstelle 10, 72076 Tübingen, Germany
[6] Physikalisches Institut, Universität Bern, Gesellschaftsstr. 6, 3012 Bern, Switzerland
[7] Univ. Grenoble Alpes, IPAG, Grenoble, France

**ABSTRACT**

*Context.* Intrinsic H$\alpha$ emission can be advantageously used to detect substellar companions because it improves contrasts in direct imaging. Characterising this emission from accreting exoplanets allows for the testing of planet formation theories.
*Aims.* We characterise the young circumbinary planetary mass companion 2MASS J01033563-5515561 (AB)b (Delorme 1 (AB)b) through medium-resolution spectroscopy.
*Methods.* We used the new narrow-field mode for the MUSE integral-field spectrograph, located on the ESO Very Large Telescope, during science verification time to obtain optical spectra of Delorme 1 (AB)b.
*Results.* We report the discovery of very strong H$\alpha$ and H$\beta$ emission, accompanied by He I emission. This is consistent with an active accretion scenario. We provide accretion rate estimates obtained from several independent methods and find a likely mass of 12−15 $M_{\rm Jup}$ for Delorme 1 (AB)b. This is also consistent with previous estimates.
*Conclusions.* Signs of active accretion in the Delorme 1 system might indicate a younger age than the ∼30−40 Myr expected from a likely membership in Tucana-Horologium (THA). Previous works have also shown the central binary to be overluminous, which gives further indication of a younger age. However, recent discoveries of active discs in relatively old (∼40 Myr), very low-mass systems suggests that ongoing accretion in Delorme 1 (AB)b might not require in and of itself that the system is younger than the age implied by its THA membership.

**Key words.** planets and satellites: individual: 2MASS J01033563-5515561 (AB)b (Delorme 1 (AB)b) – planetary systems – accretion, accretion disks – stars: low-mass – techniques: imaging spectroscopy

## 1. Introduction

As young stellar objects accrete gas as part of their formation (e.g. Hartmann et al. 2016), the heated gas produces emission lines that constitute the main observational probe of this event (e.g. Muzerolle et al. 1998). Hydrogen, which is the main constituent of the accreted gas, produces the most prominent features. While hydrogen emission lines can also result from chromospheric activity (e.g. Manara et al. 2017), sufficiently strong and wide lines are thought to result exclusively from accretion (Jayawardhana et al. 2003; Natta et al. 2004; Rigliaco et al. 2012). In analogy to this process in young stars, young brown dwarfs (BDs) and planetary mass objects have been increasingly examined for accretion signatures (e.g. Bowler et al. 2011; Joergens et al. 2013; Zhou et al. 2014; Manara et al. 2015; Petrus et al. 2020). With the advent of increasingly sophisticated adaptive optics (AO) systems, such studies can now be performed even for BDs and planetary companions at close separations from their parent stars (e.g. Sallum et al. 2015; Cugno et al. 2019). One prominent example is the PDS 70 system, where two accreting planet candidates have been observed both photospherically in the near-infrared (Keppler et al. 2018; Mesa et al. 2019) and through hydrogen emission (Haffert et al. 2019). In both cases there is also additional evidence for the circumplanetary discs from which the accreting gas is thought to originate (Christiaens et al. 2019; Isella et al. 2019).

In 2013, Delorme et al. (2013) published the discovery of a 12−14 $M_{\rm Jup}$ circumbinary companion in the system 2MASS J01033563-5515561. Based on the detection of the system architecture, the system is registered in the Washington Double Star catalog (WDS), Mason et al. (2001), as "** DLR 1". Following the example of common nomenclature for the BD binary "Luhman 16" (Mamajek 2013) based on its WDS identifier, and to avoid the cumbersome identifier in 2MASS and similar catalogues, we therefore refer to the system henceforth as "Delorme 1". The stellar components are referred to as "A" and "B", respectively, and the circumstellar companion is referred to as "(AB)b". In this Letter, we present the discovery of hydrogen and helium emission lines from Delorme 1 (AB)b, using the Integral Field Spectrograph (IFS) of the Multi Unit Spectroscopic Explorer (MUSE; Bacon et al. 2010) in its newly commissioned narrow-field mode (NFM). The lines probably indicate ongoing accretion.

## 2. System characteristics and observations

Here we present the known characteristics of the Delorme 1 system, and introduce the MUSE NFM alongside the fully resolved system. The observing log and a description of the data reduction process are available in Appendix A.1.

### 2.1. Characteristics of the Delorme 1 system

The Delorme 1 system was resolved in 2012 by Delorme et al. (2013) in the $L'$ band, revealing three components. They identified Delorme 1 AB as a ∼0.25″ (12 au) low-mass binary of spectral type (SpT) M5.5, with isochronal masses of 0.19 $M_\odot$ and 0.17 $M_\odot$ for A and B, respectively, at a distance of 47.2 ± 3.1 pc (Riedel et al. 2014). Delorme 1 (AB)b was discovered at a projected separation of ∼1.77″ (84 au) as a 12−14 $M_{Jup}$ substellar companion, assuming a system age of 30 Myr. This young age is based on a high probability of membership in the Tucana-Horologium (THA) association (Gagné et al. 2015). Riedel et al. (2014) noted that the central binary pair is over-luminous, possibly indicating an even younger age for the system (more similar to $\beta$ Pic), or further multiplicity in both components. The very red $JHK_S$ colours of Delorme 1 (AB)b are also compatible with a young planetary-mass object of an early-L SpT, providing an independent indication of youth. Using archival NACO data, Delorme et al. (2013) confirmed it to be a bound companion, and it has subsequently also been imaged in $z'$ using AstraLux by Janson et al. (2017). Based on the location of Delorme 1(AB)b in a $J-K$ colour magnitude diagram, Liu et al. (2016) found that the companion likely has very low surface gravity (VL-G; another indication of youth). It is very similar in colour to several VL-G field BDs and young companions, for example the 13−14 $M_{Jup}$ AB Pic b (SpT L0; Chauvin et al. 2005) and the ∼31 $M_{Jup}$ CD-35 2722 b (SpT L4; Wahhaj et al. 2011). Finally, the system has been studied by Binks & Jeffries (2017) using WISE data. The authors searched for an infrared excess of the unresolved stellar system that might indicate a disc. They found no clear detection of excess, but ultimately rejected the analysis because of Delorme 1 (AB)b.

### 2.2. MUSE in narrow-field mode

The observations were carried out with the MUSE IFS, located at UT4 of the European Southern Observatory (ESO) Very Large Telescope (VLT), in the new NFM, using science verification (SV) time (ID: 60.A-9485(A)). A total of 12 data cubes were obtained over 2.5 h. The IFS of MUSE covers a spectral range of 4650−9350 Å, with a resolving power R of 2000−4000 (Bacon et al. 2010). The NFM uses a new laser tomography adaptive optics (LTAO) system to provide near diffraction limited observations over a 7.5″ × 7.5″ field of view (FOV) and spatial sampling of 0.025″ × 0.025″. This increases the spatial resolution by a factor 8 over the wide-field mode. Figure 1 shows the system as resolved by MUSE.

## 3. Analysis and results

We detect very strong H$\alpha$ and H$\beta$ emission from Delorme 1 (AB)b, accompanied by emission in three He I lines (Fig. 2).

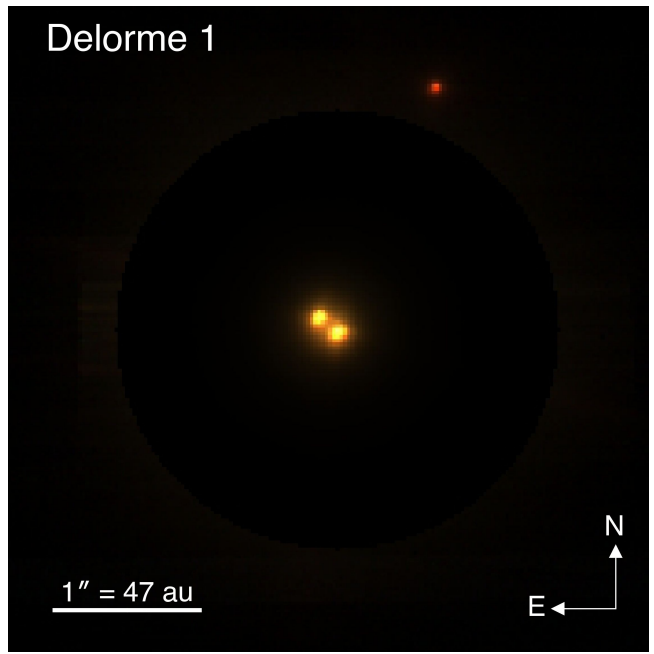

**Fig. 1.** Composite-colour image of the Delorme 1 system, created from the MUSE NFM observations. The slightly brighter primary component A is located to the south-west in the image, with B to the north-east and the significantly redder companion Delorme 1 (AB)b to the north. The central binary is scaled down in flux by a factor of 200.

This likely indicates that accretion onto this planetary-mass object is ongoing. For key lines (Table 1) we provide line fluxes and line profile widths based on which we estimate the accretion rate using three different models (Table 2). Supplementary figures and tables can be found in Appendix B.

### 3.1. Detection of very strong line emission in Delorme 1 (AB)b

The detected H$\alpha$ emission in the spectrum of Delorme 1 (AB)b (Fig. 2) clearly stands out, with an integrated flux of 1.28 ± 0.70 × $10^{-15}$ erg cm$^{-2}$ s$^{-1}$. H$\beta$ does not have a well-defined local continuum for comparison, and peaks at ∼10% of H$\alpha$. Three weak but clearly visible helium emission lines are also present at (in-air wavelengths) 6678.1517 Å, 7065.1771 Å, and 7281.35 Å, in decreasing line strength. We detect no lines in the Ca infrared triplet (IRT), and while these are also possible indicators of accretion (e.g. White & Basri 2003), they are not always observed in accreting objects (e.g. Mohanty et al. 2005). The average H$\alpha$ equivalent width (EW) of Delorme 1 (AB)b is −135 Å, which is comparable to that of other BDs that are possible accretors (e.g. Jayawardhana et al. 2003; Natta et al. 2004; Herczeg et al. 2009; Joergens et al. 2013). Extinction-corrected line fluxes, line luminosities, EWs, and line profile widths are available in Table 1.

Based on the spectrum of Delorme 1 (AB)b, and using qualitative criteria from Kirkpatrick et al. (1999), we tentatively classify Delorme 1 (AB)b as very late-M or early-L, most likely SpT L0 due to its similarity to the L0 standard BD 2MASP J0345+2540. The Rb I and Cs I doublets are weak but present, and the depth of TiO (8432 Å) is similar to that of CrH (8611 Å) and FeH (8692 Å). Stronger-than-expected absorption depth in the VO molecular bands, particularly at 7800−8000 Å, might

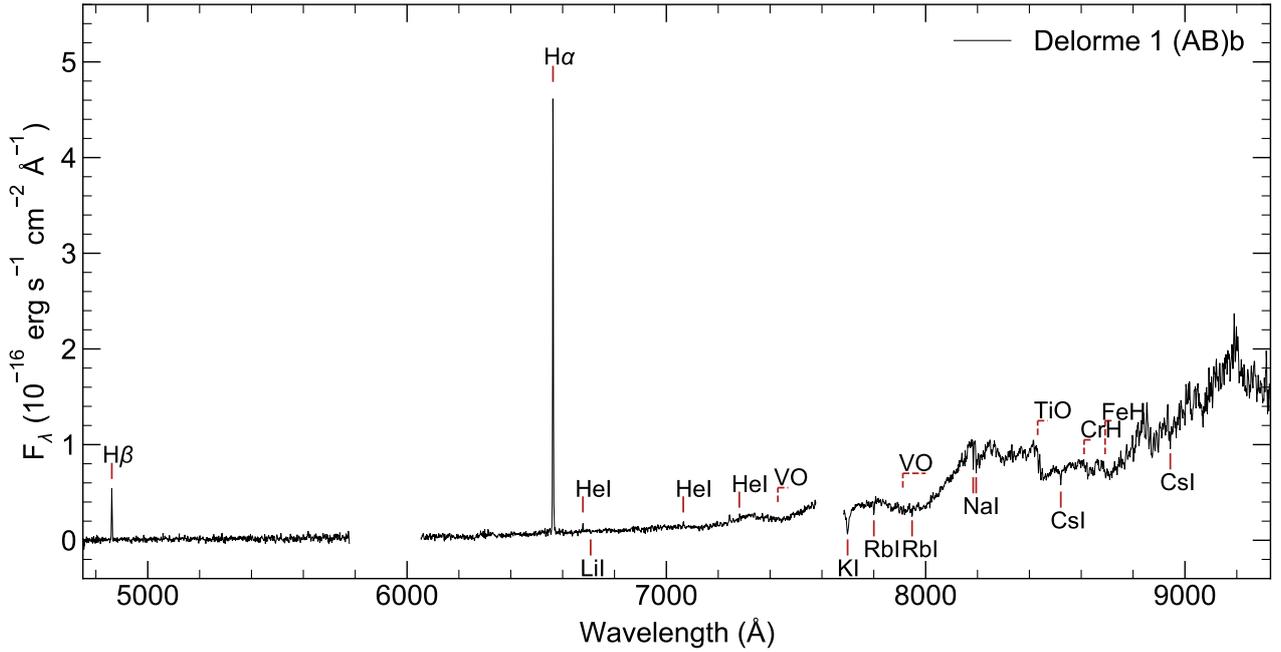

**Fig. 2.** Full averaged (over 11 cubes) MUSE spectrum of the companion Delorme 1 (AB)b (black line) showing very strong Hα emission, and highlighting other atomic lines (black text and red lines) and molecular absorption features (dashed red lines). The two narrow gaps in the spectrum are caused by sections of unusable data at 5800–6051 Å and 7592–7681 Å.

**Table 1.** Individual line characteristics (extinction-corrected flux, EW, and the 10% and 50% line profile widths) of Delorme 1 (AB)b for detected emission lines, obtained from the averaged spectrum.

| Line | $F_{\text{line}}$ ($10^{-16}$ erg cm$^{-2}$ s$^{-1}$) | EW (Å) | 10% width (km s$^{-1}$) | $FWHM$ (km s$^{-1}$) | $\Delta m_A$ (mag) | $\Delta m_B$ (mag) |
|---|---|---|---|---|---|---|
| Hβ | 1.39 ± 0.10 | … | <322 | <177 | 3.21 ± 0.23 | 3.03 ± 0.24 |
| Hα | 12.80 ± 0.70 | −135 ± 5 | 105−241 | 58−133 | 4.31 ± 0.16 | 3.89 ± 0.16 |
| He I $_{\lambda 6678}$ | 0.18 ± 0.03 | −1.9 ± 0.2 | <194 | <108 | … | … |
| He I $_{\lambda 7065}$ | 0.15 ± 0.04 | −1.0 ± 0.1 | … | … | … | … |

**Notes.** Negative EW indicates emission. For Hβ there is no well-defined continuum, required for an EW estimate, and the measured FWHM is equivalent to the resolution width of MUSE ($R_{H\beta}$ = 1694 ∼ 177 km s$^{-1}$). For HeI$_{7065}$, the line profile is not well defined enough for an accurate estimate of the width. The $\Delta m$ values compare the Hα and Hβ line fluxes of Delorme 1 (AB)b with those of the primary components A and B at a 2.5 pixel aperture.

indicate a lower-gravity atmosphere (e.g. McGovern et al. 2004), which would be in line with the expectations of youth in this object.

It is clear that the line emission is intrinsic to Delorme 1 (AB)b. We therefore turn our attention to the question of the mechanism behind it.

### 3.2. Mass accretion rates

For the purposes of the continued analysis of Delorme 1 (AB)b, we make use of the DUSTY00 atmosphere isochronal models for very-low-mass stars and BDs from Chabrier et al. (2000) and Baraffe et al. (2002) in lieu of a more in-depth spectral modelling. Based on the photometry of Delorme 1 (AB)b we obtain an absolute Gunn-i magnitude of 15.7 ± 0.3. Combining this with the $JHK_S L'$ photometry from Delorme et al. (2013) and the $z'$ from Janson et al. (2017), we find very good agreement with the expected absolute $Iz'JHK_S L'$ magnitudes for the 0.012 $M_\odot$ and 0.015 $M_\odot$ isochrones at 10 Myr and 50 Myr, respectively (Table 3). This indicates a probable hot-start mass for Delorme 1 (AB)b in the range of 12−15 $M_{\text{Jup}}$, spanning the deuterium-burning limit. This is consistent with Delorme et al. (2013). The use of cold- or warm-start models would imply a higher mass (e.g. Marleau & Cumming 2014), but we restrict ourselves here to hot starts, especially because recent modelling suggests that these are more likely (Berardo et al. 2017; Marleau et al. 2019). Thus, to calculate the mass accretion rate, we adopt a companion mass of $M_P$ = 0.012 $M_\odot$ with a bolometric luminosity $\log(L_{\text{bol}}/L_\odot) = -3.58$, and take the mean radius of the two models, $R_P = 0.163 R_\odot$ (1.59 $R_{\text{Jup}}$).

We estimated the mass accretion rate $\dot{M}_{\text{acc}}$ onto Delorme 1 (AB)b with several different methods. We took two that are in common use for stars and BDs (Natta et al. 2004 and Alcalá et al. 2017), another from Thanathibodee et al. (2019), and we also applied the new gas giant accretion model of Aoyama et al. (2018) and Aoyama & Ikoma (2019).

Line luminosities can be used to estimate $\dot{M}_{\text{acc}}$ (e.g. Mohanty et al. 2005; Herczeg & Hillenbrand 2008; Rigliaco et al. 2012; Alcalá et al. 2017). Rigliaco et al. (2012) characterised the empirical correlation between the line luminosity ($L_{\text{line}}$) and

**Table 2.** Accretion luminosities and mass accretion rates for Delorme 1 (AB)b estimated from Table 1, and upper limits inferred from the non-detections of the Ca IRT.

| Line | $L_{line}$ $\log(L_{line}/L_\odot)$ | $L_{acc}$ [a] $\log(L_{acc}/L_\odot)$ | $\dot{M}_{acc}$ [a],[b] $\log(M_{Jup} \cdot yr^{-1})$ |
|---|---|---|---|
| H$\beta$ | $-8.01 \pm 0.07$ | $-6.55 \pm 0.37$ | $-9.79 \pm 0.37$ |
| H$\alpha$ | $-7.05 \pm 0.06$ | $-6.23 \pm 0.41$ | $-9.47 \pm 0.41$ |
| He I $_{\lambda 6678}$ | $-8.90 \pm 0.09$ | $-6.43 \pm 0.64$ | $-9.67 \pm 0.64$ |
| He I $_{\lambda 7065}$ | $-8.98 \pm 0.14$ | $-6.13 \pm 0.56$ | $-9.37 \pm 0.56$ |
| (Ca II $_{\lambda 8498}$) | $< -8.38$ | $< -5.70$ | $< -8.95$ |
| (Ca II $_{\lambda 8542}$) | $< -8.36$ | $< -5.67$ | $< -8.92$ |
| (Ca II $_{\lambda 8662}$) | $< -8.54$ | $< -5.65$ | $< -8.89$ |

**Notes.** The non-detection of the Ca IRT is indicated by parentheses. [a] Obtained using the relations in Alcalá et al. (2017). [b] Derived from $L_{acc}$ using $R = 0.163 R_\odot$ from the Chabrier et al. (2000) and Baraffe et al. (2002) 0.012 $M_\odot$ model at 10 Myr.

**Table 3.** Absolute magnitudes of Delorme 1 (AB)b compared with the DUSTY00 isochrones (Chabrier et al. 2000; Baraffe et al. 2002) that agree well with the photometry.

| Filter | Delorme 1 (AB)b | 0.012 $M_\odot$ 10 Myr | 0.015 $M_\odot$ 50 Myr | Ref. |
|---|---|---|---|---|
| $M_i$ | $15.7 \pm 0.3$ | 15.77 | 15.71 | 1 |
| $M_{z'}$ | 14.0 | 14.05 [a] | 14.39 [a] | 3 |
| $M_J$ | $12.1 \pm 0.3$ | 11.86 | 11.88 | 2 |
| $M_H$ | $10.9 \pm 0.2$ | 10.95 | 11.01 | 2 |
| $M_{K_S}$ | $10.3 \pm 0.2$ | 10.25 | 10.34 | 2 |
| $M_{L'}$ | $9.3 \pm 0.1$ | 9.23 | 9.33 | 2 |

**Notes.** [a] SDSS $z'$ magnitudes from the AMES-Dusty isochrones at the same mass and age (Chabrier et al. 2000; Allard et al. 2001).
**References.** (1) This work; (2) Delorme et al. (2013); (3) Janson et al. (2017).

the accretion luminosity ($L_{acc}$), and from this, estimated $\dot{M}_{acc}$. This method was further developed by Alcalá et al. (2017) and expanded to other lines. Using the relations from Alcalá et al. (2017), we calculated the accretion luminosity and mass accretion rates for H$\beta$, H$\alpha$, HeI$_{\lambda 6678}$, and HeI$_{7065}$. Table 2 shows that these all provide consistent estimates, with a H$\alpha$ derived $\dot{M}_{acc}$ of $3.4 \times 10^{-10 \pm 0.4}$ $M_{Jup}$ yr$^{-1}$. Furthermore, we estimated upper limits of the mass accretion rate from the non-detection of the Ca IRT using the same relations. We assumed an upper limit of the peak flux at 5$\sigma$ of the noise level at each line centre. From this we then obtained an integrated flux and an accretion luminosity, and the $\dot{M}_{acc}$ upper limits for the Ca IRT shown in Table 2 are consistent with a non-detection of the Ca IRT in an accretion scenario.

The line profile width can also be used to estimate $\dot{M}_{acc}$, and a width of >200 km s$^{-1}$ is commonly used to separate accreting BDs from chromospherically active BDs (e.g. Jayawardhana et al. 2003). Natta et al. (2004) used the 10% width of the H$\alpha$ line in the following empirical relation:

$$\log(\dot{M}_{acc}) = -12.89(\pm 0.3) + 9.7(\pm 0.7) \times 10^{-3} H\alpha 10\%. \quad (1)$$

We measure an H$\alpha$ 10% width for Delorme 1 (AB)b of 105–241 km s$^{-1}$. The wide range is due to the line being only marginally resolved by MUSE ($R_{H\alpha} = 2516$, $FWHM \sim 119$ km s$^{-1}$). We discuss this, and the implication for other studies of accreting planets using MUSE, in more detail in Sect. 4. From this width, we tentatively estimate $\dot{M}_{acc} = 3 \times 10^{-8^{+0.4}_{-1.4}}$ $M_{Jup}$ yr$^{-1}$ using Eq. (1), which is a factor $\sim 10^2$ greater than what we obtain from the line luminosity above.

A similar estimate can be obtained from the planet-surface shock model of Aoyama & Ikoma (2019), which combines the line luminosity with the H$\alpha$ line profile widths to estimate $\dot{M}_{acc}$, as well as the planetary mass $M_P$ (independently of evolutionary models). Using a radius $R_P = 0.163 R_\odot$, we tentatively estimate an accretion rate of $\dot{M}_{acc} = 1 \times 10^{-8}$ $M_{Jup}$ yr$^{-1}$, and a planetary mass $M_P = 11$ $M_{Jup}$. If the true line profile widths are closer to the lower limits of the width, this would indicate a similar $\dot{M}_{acc}$ but a substantially lower mass, down to $M_P \sim 6$ $M_{Jup}$. We note that this lower estimate is discrepant from the estimate provided by the evolutionary models (Table 3). The accretion rates derived from the methods of Natta et al. (2004) and Aoyama & Ikoma (2019) are consistent with the $\dot{M}_{acc} = 0.8 \times 10^{-8}$ $M_{Jup}$ yr$^{-1}$ obtained for Delorme 1 (AB)b using the H$\alpha$ line luminosity based model of Thanathibodee et al. (2019). It is important to note that the applicability of this method to the case of Delorme 1 (AB)b depends on it having a comparable mass and radius to the PDS 70 planets (Thanathibodee et al. 2019). While both parameters are still highly uncertain for these planets, the mass estimates for Delorme 1 (AB)b thus far are comparable to PDS 70 b, and to the radius and mass assumed for the model by Thanathibodee et al. (2019) ($M_P = 6$ $M_{Jup}$, $R_P = 1.3$ $R_{Jup}$).

## 4. Discussion

We have presented photometry and spectroscopy from MUSE NFM observations of the Delorme 1 system. From the analysis of the detected emission lines (Fig. 2), we found several indicators that accretion is ongoing in Delorme 1 (AB)b. Using four different methods, we estimated mass accretion rates $\dot{M}_{acc}$ of $0.8-3.0 \times 10^{-8}$ $M_{Jup}$ yr$^{-1}$ (from Natta et al. 2004; Aoyama & Ikoma 2019; Thanathibodee et al. 2019) and $3.4 \times 10^{-10}$ $M_{Jup}$ yr$^{-1}$ (from Alcalá et al. 2017). In Appendix A.2 we also investigate the potential emission contribution by chromospheric activity, which is likely negligible. Based on these results, there are strong indications that Delorme 1 (AB)b is actively accreting. We now aim to discuss this in more detail, and conclude with a brief discussion on the age of the Delorme 1 system. We also determine whether it excludes the hypothesis of ongoing accretion in Delorme 1 (AB)b.

The limited resolution of MUSE affects the precision of $\dot{M}_{acc}$ estimates that make use of line profile shape measurements, such as Eq. (1) (Natta et al. 2004) and in part the model by Aoyama & Ikoma (2019) (which is still constrained by the H$\alpha$ line luminosity). The H$\beta$ line FWHM is essentially equal to the line-broadening FWHM of MUSE at 4861 Å (177 km s$^{-1}$), so it is unresolved and cannot be used to distinguish the intrinsic line shape. The H$\alpha$ line, on the other hand, has a somewhat larger FWHM (133 km s$^{-1}$) than the line-broadening profile at 6563 Å (119 km s$^{-1}$), therefore it is marginally but not fully resolved. The reported H$\alpha$ FWHMs for PDS 70 b and c from Haffert et al. (2019) are smaller than Delorme 1 (AB)b and are effectively unresolved, with widths similar to the FWHM of the line-broadening profile (see also Thanathibodee et al. 2019; Hashimoto et al. 2020), which means that they cannot be meaningfully compared with Delorme 1 (AB)b in this context. While there is clear evidence that the PDS 70 (proto)planets are accreting, this suggests that as MUSE becomes an ever more important instrument in the hunt for accreting planets (Girard et al. 2020), its limitations should be kept in mind as well.

As mentioned in Sect. 3.2, there is a factor $\sim 10^2$ difference in the mass accretion rates estimated for Delorme 1 (AB)b by the results of Natta et al. (2004), Aoyama & Ikoma (2019) and Thanathibodee et al. (2019) on the one hand, and Alcalá et al. (2017) on the other. This discrepancy could be related to the uncertainty on the line profiles and line ratios, or to that on the origin of the emission (and potential absorption thereof). For example, in the stellar case, hydrogen is excited in the pre-shock gas and is ionised post-shock, while for planetary mass objects, the pre-shock temperatures are too low for the excitation of hydrogen (Aoyama & Ikoma 2019). Following the accretion shock, however, temperatures are high enough to produce hydrogen emission.

In Sect. 2.1 we noted that the age of the Delorme 1 system is still not well constrained. Several works assign Delorme 1 to the THA (e.g. Kraus et al. 2014; Malo et al. 2014; Gagné et al. 2015), while Riedel et al. (2014) noted that the over-luminosity of both Delorme 1 A and B indicates an even younger age. Ujjwal et al. (2020) used data from *Gaia* DR2 and found a much greater age span for THA, 3–60.5 Myr, with an estimated average age of ~20 Myr, and several other moving groups. This might add uncertainty to the true age of THA. It is worth noting that an age for Delorme 1 of 30–45 Myr does not automatically exclude it as a possible site of accretion. Murphy et al. (2018) identified the 45 Myr old 0.1 $M_\odot$ M5 WISE J0808-6443 as the host of a primordial disc, and these authors were recently joined by Lee et al. (2020), who reported a likely transitional disc around another 0.1 $M_\odot$, M5 star, the 55 Myr old 2MASS J15460752-6258042. Silverberg et al. (2020) identified another four, coining the term "Peter Pan discs" for these types of long-lived accretion discs that continue to be found around very low-mass stars. Perhaps even more interestingly, Boucher et al. (2016) reported the detection of a disc around the L0, ~14 $M_{Jup}$ BD and fellow THA member 2MASS J02265658-5327032. Taken together with our results, we find that an in-depth investigation of the Delorme 1 system is well motivated.

To conclude, the implied masses of the three components indicate a high ratio of companion to host star/system mass of ~0.03−0.04. This makes it difficult to constrain possible formation scenarios for Delorme 1 (AB)b (Delorme et al. 2013). In light of this, the possibility that the companion Delorme 1 (AB)b is still actively accreting could be an important piece to the puzzle. For example, higher resolution observations of the emission lines could provide more precise accretion estimates and allow for comparison with different classes of very low-mass objects, such as the populations studied in Zhou et al. (2014) and Manara et al. (2015), respectively.

*Acknowledgements.* We thank the anonymous referee for the very useful comments that improved this Letter. R. Asensio-Torres acknowledges support from the European Research Council under the Horizon 2020 Framework Program via the ERC Advanced Grant Origins 83 24 28. MJ gratefully acknowledges support from the Knut and Alice Wallenberg Foundation. G-DM acknowledges the support of the DFG priority program SPP 1992 "Exploring the Diversity of Extrasolar Planets" (KU 2849/7-1) and the support from the Swiss National Science Foundation under grant BSSGI0_155816 "PlanetsInTime". Parts of this work have been carried out within the framework of the NCCR PlanetS supported by the Swiss National Science Foundation. This research has made use of the VizieR catalogue access tool, CDS, Strasbourg, France (Ochsenbein et al. 2000), the Washington Double Star Catalog maintained at the US Naval Observatory, and Photutils, an Astropy package for detection and photometry of astronomical sources (Bradley et al. 2019).

## Appendix A: MUSE data reduction and chromospheric activity

*A.1. MUSE data reduction*

**Table A.1.** Observing log of the 2018 September 18 observation of Delorme 1 under ESO SV programme ID 60.A-9485(A).

| UT time (hh:mm) | Seeing ($''$) | Airmass |
|---|---|---|
| 02:55 [a] | 0.72 [a] | 1.41 [a] |
| 03:03 | 0.71 | 1.38 |
| 03:10 | 0.66 | 1.36 |
| 03:18 | 0.61 | 1.34 |
| 03:25 | 0.67 | 1.32 |
| 03:33 | 0.76 | 1.30 |
| 04:57 | 0.56 | 1.18 |
| 05:05 | 0.53 | 1.18 |
| 05:12 | 0.55 | 1.17 |
| 05:21 | 0.66 | 1.17 |
| 05:28 | 0.66 | 1.17 |

**Notes.** (DIT, NDIT) of (340s, 1) for all integrations. [a] Excluded from the analysis of Delorme 1 (AB)b due to very poor quality photometry of the companion.

Table A.1 lists the 12 integrations obtained over 2.5 h during overall favourable conditions. These were executed using the NFM generic offset observing template, which includes a small-step dither of $0.05''$ (2 pixels) and a field rotation of $90°$ between each exposure in a four-point pattern.

We used the MUSE data reduction pipeline (Weilbacher et al. 2012) to process the observations. Each observation resulted in a data cube consisting of 3861 individual frames. The wide angular separation between the central binary and the companion, combined with the high-performance AO, results in the companion being easily resolved, as shown in Fig. 1. It is located well outside the halo of the stellar point spread function (PSF), which means that stellar subtraction techniques are not required.

Aperture photometry using the Python routine photutils was then used to extract the spectrum for Delorme 1 (AB)b. Circular aperture radii in the range of 2.0–6.0 pixels were used in increments of 0.5 pixels. Background was subtracted by placing two circular apertures of the same radius on either side of the target, equidistant from the photometric centre, providing an average value of the background at the target position. An aperture radius of 4.0 pixels was found to provide a good balance between a high signal-to-noise ratio (S/N) and good coverage of the companion PSF. A similar technique was used for Delorme 1 A and B, with an aperture radius of 2.5 pixels.

Flux was calibrated automatically in the pipeline, and was later verified by integrating the extinction-corrected ($A_V$ = 0.0582; Schlafly & Finkbeiner 2011) flux from the central pair over a suitable band pass. For individual line fluxes, the line-specific extinction correction was calculated from Cardelli et al. (1989), and this value was used for $A_V$. Delorme 1 was observed in the Gunn-$i$ (Oke & Gunn 1983) band during the DEep Near-Infrared Survey (DENIS; DENIS Consortium 2005) at 11.38 mag. Using a wide aperture for the central pair and a four-pixel aperture for the companion, we obtain apparent Gunn-$i$ magnitudes of $11.31 \pm 0.06$ mag for Delorme 1 AB and $19.08 \pm 0.16$ mag for Delorme 1 (AB)b. This corresponds to an absolute magnitude of $15.7 \pm 0.3$, assuming a distance of $47.2 \pm 3.1$ pc. This brightness is comparable to BDs of similar SpT and distance as Delorme 1 (AB)b.

*A.2. Chromospheric activity*

An alternative to the hypothesis that Delorme 1 (AB)b is actively accreting is that the emission is instead due to chromospheric activity. H$\alpha$ emission thought to be originating from chromospheric activity has been observed in a number of BDs. Recent examples include the more massive BD companion PZ TEL B (Musso Barcucci et al. 2019) and a survey of Sco-Cen BDs by Petrus et al. (2020), but typically with lower levels of emission than we find for Delorme 1 (AB)b. Manara et al. (2013) obtained an analytical relation from observations of a number of young (<10 Myr) K7 – M9.5 stars and BDs for the purpose of estimating the chromospheric activity contribution as an accretion-luminosity noise. This work was extended by Manara et al. (2017) by including more stars in the lowest mass-range, resulting in the relation

$$\log(L_{\rm acc,noise}/L_{\rm bol}) = (6.2 \pm 0.5) \cdot \log T_{\rm eff} - (24.5 \pm 1.9). \quad (A.1)$$

While this relation includes substantial uncertainties and Delorme 1 (AB)b may fall outside it because it is a very low-mass object, it is worth investigating the potential contribution from chromospheric activity. The 10 Myr, $0.012\,M_\odot$ isochrone (Sect. 3.2) yields an effective temperature $T_{\rm eff}$ = 1801 K and $\log(L_{\rm bol}/L_\odot) = -3.58$, resulting in $\log(L_{\rm acc,noise}/L_\odot) = -8.0 \pm 2.6$. Based on the relation from Alcalá et al. (2017), this implies a mass accretion rate noise $\dot{M}_{\rm acc,noise} = 5.2 \times 10^{-12\pm2.6}\,M_{\rm Jup}\,{\rm yr}^{-1}$, which is a factor $10^{-2}$–$10^{-4}$ lower than the previously estimated accretion rates in this work. Activity is generally expected to decrease with age (e.g. Zuckerman & Song 2004; Soderblom 2010; Malo et al. 2014). Consequently, from the point of view of activity, it should be even less likely to measure such high H$\alpha$ emission in Delorme 1 (AB)b at ages later than 10 Myr.

## Appendix B: Supplementary material

This appendix includes a number of supplementary figures and tables. Figure B.1 shows the spectra of the central binary pair, with both components showing signs of strong hydrogen emission. The nature of this emission will be explored further in Eriksson et al. (in prep.). By comparing the spectrum of Delorme 1 (AB)b with the spectra of the central pair (Fig. B.1), we can immediately note a number of differences that indicate that detected spectral features and line emission are intrinsic to each object. (i) The primary component, A, is undergoing a flare event at the time of observation. This is evident from comparing the temporal evolution of the EW during the observation (Table B.1 and Fig. B.2), which for Delorme 1 (AB)b shows an evolution that does not correlate with either A or B. (ii) He I emission can only be seen in the companion, and not in the primary or secondary component. (iii) The H$\alpha$/H$\beta$ line flux ratios of A and B are ~1/3 of (AB)b, and the 10% and 50% line profile widths of H$\alpha$ and H$\beta$ are narrower than the stellar counterparts.

Table B.2 lists the line characteristics for Delorme 1 (AB)b (H$\beta$ and H$\alpha$) for the individual datacubes, with accompanying mass accretion rates listed in Table B.3. Figure B.3 shows the resolution curve for MUSE over its wavelength coverage. The interpolated R values for the lines are discussed in Sect. 3, along with the corresponding line profile $\Delta v$. Finally, Fig. B.4

illustrates to which extent the Hβ and Hα line profiles are solved by MUSE in Delorme 1 (AB)b, and Fig. B.5 shows that the line profiles are better resolved for Delorme 1 A and B than for Delorme 1 (AB)b. To estimate how well Hα is resolved, we convolved the MUSE line-broadening profile with continuously narrower widths and locate the smallest width that would produce a convolution matching the observed profile. To illustrate how the uncertainty is reduced at greater widths, we repeated this estimation for Delorme 1 A and find a 10% width of $217-307 \text{ km s}^{-1}$ (FWHM $169 \text{ km s}^{-1}$).

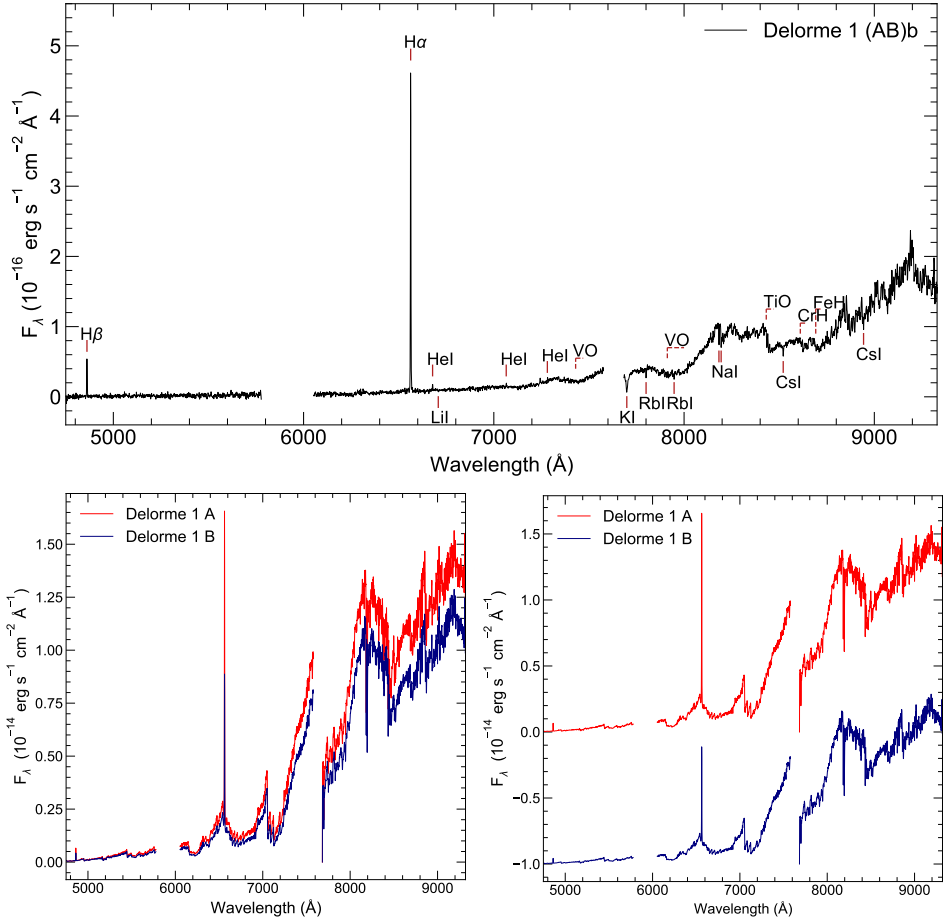

**Fig. B.1.** *Upper panel*: spectrum of Delorme 1 (AB)b as shown in Fig. 2. *Bottom left*: full averaged (over 12 cubes) MUSE IFS spectrum of the two components of the central binary pair, Delorme 1 AB, indicating nearly identical spectral types (~M5−M6). *Bottom right*: secondary component (B) shifted down in flux by $10^{-14}$ for clarity. Prominent features include very strong Hα and Na I doublet, and weaker Hβ lines in both components. The primary component (A) shows signs of Ca IRT emission, with the first two Ca II lines visible at 8498/8542 Å. This is likely associated with the recent flare because it is not detectable in B. We also note the lack of any He I features in the 6678−7281 Å range, in contrast to the companion spectrum. We detect a blueshift of $\sim -15 \text{ km s}^{-1}$ in the Hα line of Delorme 1 A during the flare event, which is not detected in B. We detect no significant blue- or redshift in the lines of Delorme 1 (AB)b.

**Table B.1.** Hα EWs of the three components, A, B, and (AB)b, in the Delorme 1 system during the observation.

| UT time (hh:mm) | $EW_{A,H\alpha}$ (Å) | $EW_{B,H\alpha}$ (Å) | $EW_{(AB)b,H\alpha}$ (Å) |
|---|---|---|---|
| 02:55 | $-68.6 \pm 0.3$ | $-32.1 \pm 0.1$ | ... [a] |
| 03:03 | $-60.3 \pm 0.1$ | $-29.7 \pm 0.1$ | $-112 \pm 10$ |
| 03:10 | $-49.9 \pm 0.0$ | $-25.0 \pm 0.5$ | $-99 \pm 11$ |
| 03:18 | $-49.5 \pm 0.4$ | $-31.6 \pm 0.2$ | $-166 \pm 8$ |
| 03:25 | $-46.1 \pm 0.3$ | $-32.0 \pm 0.5$ | $-147 \pm 5$ |
| 03:33 | $-44.7 \pm 0.3$ | $-33.7 \pm 0.1$ | $-120 \pm 9$ |
| 04:57 | $-38.1 \pm 0.3$ | $-30.6 \pm 1.0$ | $-200 \pm 21$ |
| 05:05 | $-37.9 \pm 0.1$ | $-29.4 \pm 0.1$ | $-156 \pm 35$ |
| 05:12 | $-37.2 \pm 0.4$ | $-33.4 \pm 0.2$ | $-118 \pm 11$ |
| 05:21 | $-39.3 \pm 0.5$ | $-32.7 \pm 0.0$ | $-124 \pm 5$ |
| 05:28 | $-41.8 \pm 0.1$ | $-31.8 \pm 0.4$ | $-141 \pm 16$ |
| 05:36 | $-38.5 \pm 0.1$ | $-34.8 \pm 0.3$ | $-157 \pm 6$ |

**Notes.** Negative values indicate emission. See Fig. B.2 for a graphical representation of this table. [a] Missing value because this cube was excluded from the analysis of the companion (Sect. 2).

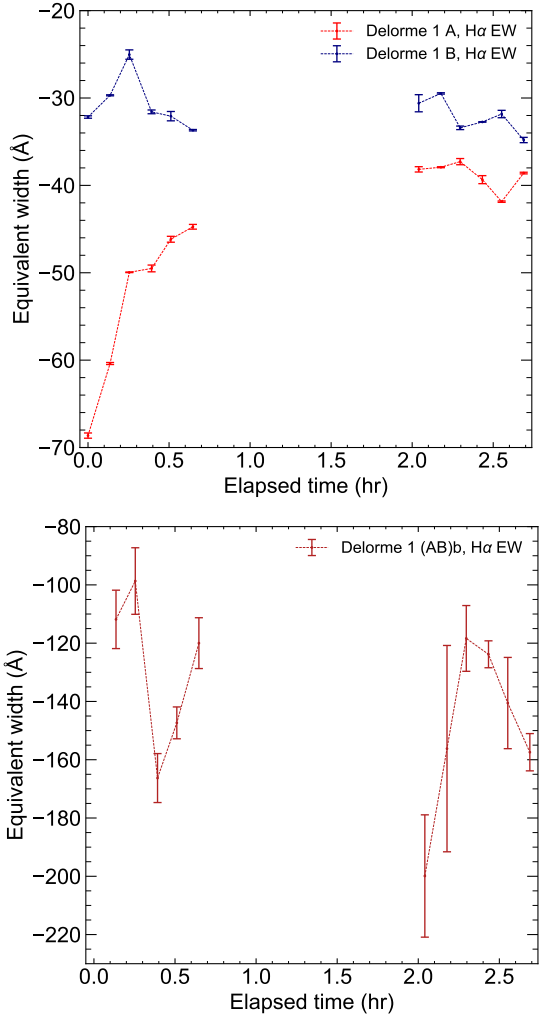

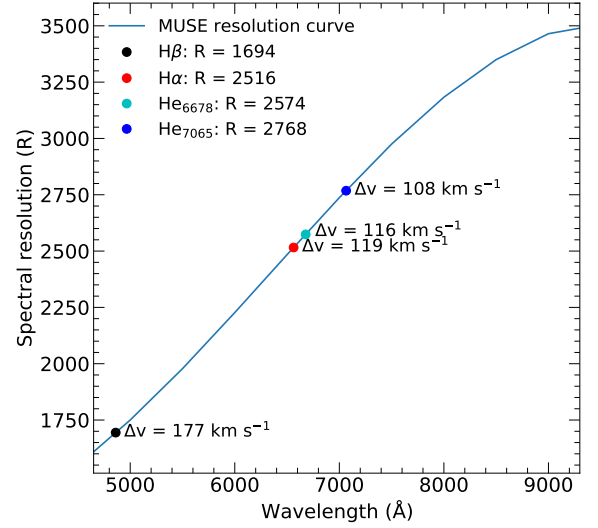

**Fig. B.3.** Interpolated values for R at the four emission lines used in the data analysis (coloured points). The blue line is obtained from the values listed in 500 Å increments, in the MUSE User Manual (version 11.3, Fig. 18). The $\Delta v$ values given for each point represent the FWHM of the MUSE line profile at that resolution.

**Fig. B.2.** H$\alpha$ EWs during the observation (from Table B.1) of the central binary components Delorme 1 A and Delorme 1 B (*top*), and companion Delorme 1 (AB)b (*bottom*). Negative values indicate emission. From the evolution of the EW over 2.5 h, it is clear that A is undergoing a rapid decrease in width (weaker emission) until it plateaus at around −40 Å, while B stays at an average of −32 Å, and (AB)b fluctuates around an average of −135 Å. There is no clear correlation between the changes in EW in the central binary pair and the companion, indicating that emission is intrinsic to each object. We also note that the rapid change in the EW of Delorme 1 A, as the excess emission due to the flare dissipates, is absent in B and (AB)b.

**Table B.2.** Apparent and extinction-corrected ($A_V = 0.0582$, with $A_{line}$ calculated from Cardelli et al. 1989) H$\beta$ and H$\alpha$ emission line fluxes, line luminosities, accretion luminosities, and the 10% H$\alpha$ width range during the observation on 2018 September 18.

| UT time (hh:mm) | $F_{H\beta}$ ($10^{-15}$erg.cm$^{-2}$.s$^{-1}$) | $F_{H\alpha}$ ($10^{-15}$erg.cm$^{-2}$.s$^{-1}$) | $L_{H\beta}$ log($L_{line}/L_\odot$) | $L_{H\alpha}$ log($L_{line}/L_\odot$) | $L_{acc,H\beta}$ [a] log($L_{acc}/L_\odot$) | $L_{acc,H\alpha}$ [a] log($L_{acc}/L_\odot$) | 10% H$\alpha$ (km.s$^{-1}$) |
|---|---|---|---|---|---|---|---|
| 03:03 | 0.11 ± 0.06 | 1.08 ± 0.38 | −8.08 ± 0.30 | −7.12 ± 0.20 | −6.62 ± 0.50 | −6.31 ± 0.46 | 120–248 |
| 03:10 | 0.14 ± 0.05 | 1.11 ± 0.38 | −8.00 ± 0.24 | −7.11 ± 0.20 | −6.53 ± 0.45 | −6.30 ± 0.46 | 93–236 |
| 03:18 | 0.16 ± 0.08 | 1.33 ± 0.43 | −7.93 ± 0.28 | −7.03 ± 0.18 | −6.45 ± 0.47 | −6.20 ± 0.45 | 126–251 |
| 03:25 | 0.14 ± 0.06 | 1.13 ± 0.37 | −8.00 ± 0.27 | −7.10 ± 0.19 | −6.53 ± 0.47 | −6.29 ± 0.45 | 101–240 |
| 03:33 | 0.13 ± 0.04 | 1.34 ± 0.40 | −8.04 ± 0.20 | −7.03 ± 0.16 | −6.58 ± 0.43 | −6.20 ± 0.44 | 98–238 |
| 04:57 | 0.16 ± 0.07 | 1.46 ± 0.45 | −7.94 ± 0.27 | −6.99 ± 0.17 | −6.46 ± 0.47 | −6.16 ± 0.44 | 114–245 |
| 05:05 | 0.19 ± 0.09 | 1.51 ± 0.51 | −7.87 ± 0.28 | −6.98 ± 0.19 | −6.38 ± 0.48 | −6.14 ± 0.45 | 109–243 |
| 05:12 | 0.19 ± 0.09 | 1.59 ± 0.50 | −7.86 ± 0.28 | −6.95 ± 0.18 | −6.37 ± 0.48 | −6.12 ± 0.44 | 93–236 |
| 05:21 | 0.11 ± 0.05 | 1.05 ± 0.39 | −8.09 ± 0.29 | −7.13 ± 0.22 | −6.63 ± 0.49 | −6.32 ± 0.47 | 101–240 |
| 05:28 | 0.09 ± 0.04 | 1.22 ± 0.39 | −8.17 ± 0.27 | −7.07 ± 0.18 | −6.73 ± 0.48 | −6.25 ± 0.45 | 73–229 |
| 05:36 | 0.08 ± 0.04 | 0.86 ± 0.32 | −8.22 ± 0.31 | −7.22 ± 0.22 | −6.79 ± 0.51 | −6.42 ± 0.48 | 126–251 |

**Notes.** [a] Accretion luminosity from the relations in Alcalá et al. (2017).

**Table B.3.** Resulting mass accretion rates from the accretion luminosities and 10% Hα widths in Table B.2.

| UT time (hh:mm) | $\dot{M}_{\mathrm{acc},H\beta}$ [a] $\log(M_{\mathrm{Jup}}.\mathrm{yr}^{-1})$ | $\dot{M}_{\mathrm{acc},H\alpha}$ [a] $\log(M_{\mathrm{Jup}}.\mathrm{yr}^{-1})$ | $\dot{M}_{\mathrm{acc},10\%H\alpha}$ [b] $\log(M_{\mathrm{Jup}}.\mathrm{yr}^{-1})$ |
|---|---|---|---|
| 03:03 | $-9.87 \pm 0.50$ | $-9.55 \pm 0.46$ | $-7.46^{+0.35}_{-1.29}$ |
| 03:10 | $-9.78 \pm 0.45$ | $-9.54 \pm 0.46$ | $-7.58^{+0.35}_{-1.43}$ |
| 03:18 | $-9.70 \pm 0.48$ | $-9.45 \pm 0.45$ | $-7.44^{+0.35}_{-1.26}$ |
| 03:25 | $-9.77 \pm 0.47$ | $-9.53 \pm 0.45$ | $-7.54^{+0.35}_{-1.39}$ |
| 03:33 | $-9.82 \pm 0.43$ | $-9.45 \pm 0.44$ | $-7.55^{+0.35}_{-1.40}$ |
| 04:57 | $-9.71 \pm 0.47$ | $-9.41 \pm 0.44$ | $-7.49^{+0.35}_{-1.32}$ |
| 05:05 | $-9.63 \pm 0.48$ | $-9.39 \pm 0.45$ | $-7.50^{+0.35}_{-1.35}$ |
| 05:12 | $-9.61 \pm 0.47$ | $-9.36 \pm 0.44$ | $-7.58^{+0.35}_{-1.43}$ |
| 05:21 | $-9.88 \pm 0.49$ | $-9.57 \pm 0.47$ | $-7.54^{+0.35}_{-1.39}$ |
| 05:28 | $-9.97 \pm 0.48$ | $-9.50 \pm 0.45$ | $-7.65^{+0.35}_{-1.56}$ |
| 05:36 | $-10.03 \pm 0.51$ | $-9.67 \pm 0.48$ | $-7.44^{+0.35}_{-1.26}$ |

**Notes.** [a] Mass accretion rates from the relation in Alcalá et al. (2017). [b] Mass accretion rates from Eq. (1) (Natta et al. 2004).

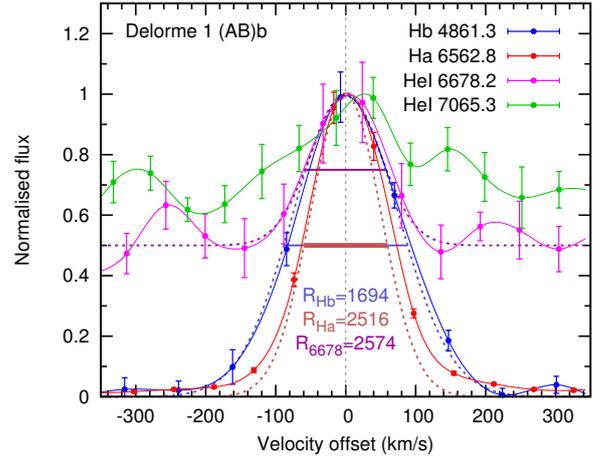

**Fig. B.4.** Comparison between observed spectral line profiles with errors (solid lines) and the line-broadening profile of MUSE at each line except for He I λ 7065 (horizontal bars at FWHM and dashed lines). The normalised (pseudo-)continuum level for He I λ 6678 is fit by eye to $f_{\min} = 0.5$. Hα is marginally resolved (wider than the dashed line-broadening profile), while Hβ is unresolved, and we therefore consider the measured Hβ line profile widths to be upper limits.

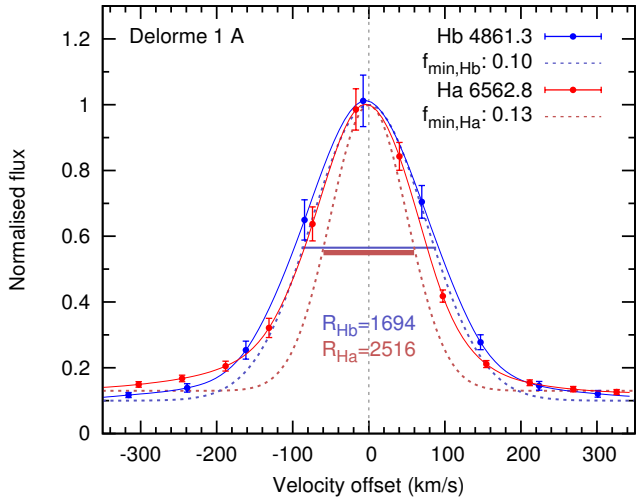
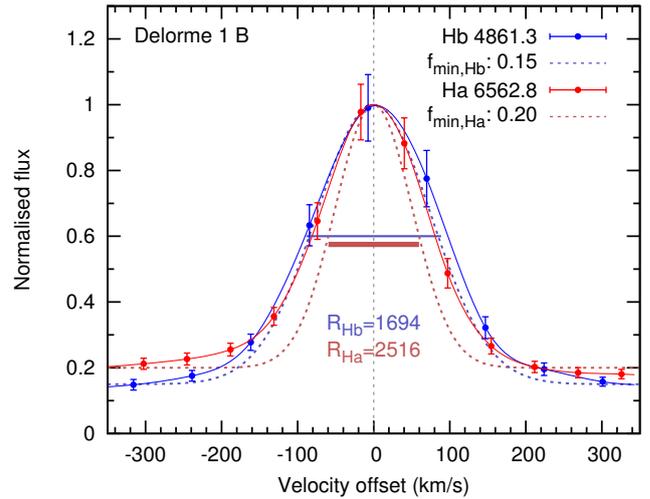

**Fig. B.5.** Similar to Fig. B.4, but for the central pair Delorme 1 A and Delorme 1 B. The $f_{\min}$ values denote the continuum level. Compared with Delorme 1 (AB)b, the Hα profiles are much better resolved here, while Hβ is marginally resolved. We estimate the Delorme 1 A 10% widths of Hβ and Hα to be 195–353 km s$^{-1}$ and 217–307 km s$^{-1}$, respectively.